\documentclass[pra,twocolumn,showpacs]{revtex4}
\usepackage{graphicx,amsmath}

\newcommand{\sinc}{\mathrm{sinc}}
\newcommand{\avg}[1]{\langle #1\rangle}

\begin{document}
\title{Intensity of parametric fluorescence pumped by ultrashort pulses}
\author{Jan Chwede\'{n}czuk}
\affiliation{Institute of Theoretical Physics, Warsaw University, Ho{\.z}a 69, 00-681 Warsaw, Poland}
\author{Wojciech Wasilewski}
\affiliation{Institute of Experimental Physics, Warsaw University, Ho{\.z}a 69, 00-681 Warsaw, Poland}
\begin{abstract}
We investigate intensity of the  parametric down conversion of ultrashort, ultraviolet pulse both in a low- and
high-conversion regime.
In the first regime, we develop a simple analytical expressions for the photon flux of the
fluorescence. Numerical simulations using three-dimensional stochastic Wigner method provide
results in the latter regime. We find that the perturbative approximation for the photon flux as a function
of the wavelength of the signal and the observation angle,
quantitatively describes even the non-perturbative regime. For short pump pulses the intensity of the fluorescence
is highest at frequencies where the group velocities of the pump and the down-conversion are equal. 
For longer pump pulses this requirement is gradually relaxed. Additionally, for small pump beams, the intensity
strongly depends on the spatial divergence between interacting fields. 
\end{abstract}
\pacs{42.65.Lm, 42.65.Yj, 42.65.Re} \maketitle
\section{Introduction}
Quantum parametric fluorescence is one of the most frequently used sources of nonclassical radiation. 
In the low gain regime, the parametric process is used as a robust source of single photons and photon pairs. 
The parametric fluorescence--based
sources allowed for a number of fundamental experiments. Examples include Bell inequality breaking 
\cite{KwiatPRL95}, quantum
cryptography \cite{GisinRMP02} and quantum teleportation \cite{Bouwmeester97}. 
Nowadays the effort focuses on engineering the correlation properties 
\cite{GricePRA01,GiovannettiPRL_88PRL,URenQIC03,TsangPRA05} of such sources.
On the other hand, in some experiments intense pump pulses are used, which results in high parametric gain.
This regime is characterized by an appreciable amount of multiple pair creation events and most
conveniently described by recalling the notion of squeezing \cite{ScullyQO}. Single or two-mode
squeezing described in textbooks can be
used to describe realistic multimode parametric amplifier \cite{WasilewskiPRA06A} and it turns out that
a laboratory device
produces a superposition of squeezed states in modes defined both by the nonlinear medium and the pump pulse.
High-gain parametric amplifiers are nowadays also frequently used for amplification of ultrashort pulses 
\cite{RiedleOL1997}. 
In this application the parametric fluorescence is an important and troublesome source of noise. 

In this paper we calculate the average angular and spectral intensity of a spontaneous parametric fluorescence.
Since parametric fluorescence has quantum origin, its intensity cannot be calculated using classical nonlinear optics.
Below, we adopt both perturbative and stochastic approach for this task. With a few approximations
we provide a simple formula for the intensity of the fluorescence in single-pair generation regime. 
Numerical simulations show, that it qualitatively reproduces main features of the donwnconversion spectrum
in the high gain regime as well. 

Our results could be used as guidelines for constructing brighter photon-pair sources. Other
application would be optimization of high-gain parametric amplifiers for a minimal contribution of 
parametric fluorescence \cite{TavellaNJP06}. 

Let us mention that a process analogous to parametric down-conversion
is observed in physics of cold atoms. When a pair Bose-Einstein condensates
collide, a halo of spontaneously scattered of atoms is observed \cite{VogelsPRL02,PerrinPRL07}. Stochastic methods used for
description of this process proved extremely successful \cite{NorriePRL05,DeuarPRL07} and we adopted them 
to obtain results presented below. 

The paper is organized as follows. In Section II we recall
some basics of the theory of the parametric down-conversion. In Section III, we calculate the photon flux in the
perturbative regime. In Section IV, we compare these results with numerical predictions in the high-gain regime. 
We conclude in Section V. 

\section{Theory of the parametric down-conversion}
\newcommand{\LNL}{L_\mathrm{NL}}
\newcommand{\inp}{_\mathrm{in}}
\newcommand{\out}{_\mathrm{out}}
\renewcommand{\o}{\omega}
\newcommand{\V}[1]{\mathbf{#1}}
\newcommand{\kp}{\V{k}_\perp}
\newcommand{\rp}{\V{r}_\perp}

A process of parametric down-conversion takes place in a crystal with non-zero second order nonlinear susceptibility
tensor $\chi^{(2)}$ when it is irradiated with a pump beam. During pass through the nonlinear medium each 
of the pump pulses
gives rise to photon pairs which form the parametric fluorescence. Below, we formulate a theoretical model 
describing the nonlinear interaction which allows for calculation of the fluorescence intensity.

\begin{figure}
 	\centering
	\includegraphics[scale=0.7]{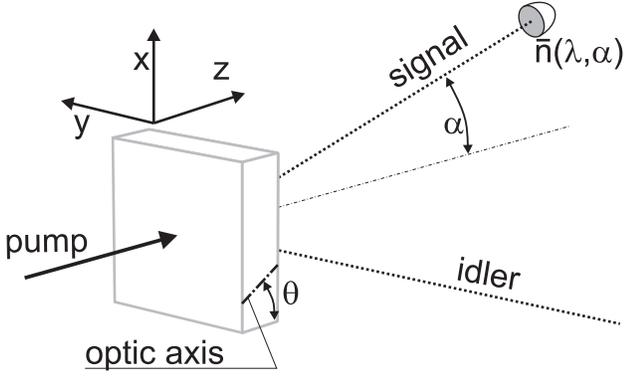}
 	\caption{Scheme of a measurement of the photon flux $\bar{n}(\lambda,\alpha)$ 
 	of the spontaneous parametric fluorescence. The nonlinear crystal is pumped along $z$-axis. 
 	Out of the emerging parametric fluorescence we observe
 	a small portion at the wavelength $\lambda$ propagating at an angle $\alpha$ to $z$-axis in the $x$-$z$ plane.}
 	\label{fig:SPDC}
\end{figure}
We assume, that the medium interfaces are parallel to
$x$--$y$ plane, while fields propagate along the $z$-coordinate, as depicted in Fig.~\ref{fig:SPDC}. 
We expand all the fields involved in the
interaction using an approach from nonlinear optics. At each $z$ an expansion in a basis of plane waves with frequency
$\omega$ and wave-vector $k_x$, $k_y$ is given, while $z$ parametrizes the development of the evolution.
To shorten the notation, let's denote the triple $(\omega,k_x,k_y)$ parametrizing each plane-wave component 
of the fields by $\kappa$.
For concreteness, we consider Type-I interaction in BBO ($\beta$-Barium Borate) in which pump propagates as an 
$e$-ray while
the fluorescence as an $o$-ray. Results can however be easily generalized for any nonlinear crystal by substituting its
particular dispersion relation in the formulas below.

Let's consider a regime in which the pump consists of intense pulses described by a classical envelope
$\tilde A_p(\kappa_p,z)$ in the frequency domain, spanned by $\kappa_p=(\omega_p,k_{px},k_{py})$.
The evolution of the pump pulse $\tilde A_p(\kappa_p,z)$ along $z$ is determined by the dispersion relation of 
the crystal.
A nonlinear interaction with the fluorescence field can be neglected, since 
the intensity of the pumping field is large and its depletion due to parametric down conversion is minimal.
This condition is readily satisfied in many experiments. Note that the intensity of the pump is also assumed large
enough to neglect any quantum fluctuations and describe it by a classical field.

For a pump propagating as an $e$-ray in a crystal cut at an angle $\theta$ to the $z$-axis the dispersion relation 
of the crystal reads:
\begin{multline}
\frac{\omega_p^2}{c^2}=
\frac{(k_{px}\cos\theta-k_{pz}\sin\theta)^2}{n_e^2(\omega_p)} \\
+\frac{(k_{pz}\cos\theta+k_{px}\sin\theta)^2+k_{py}^2}{n_o^2(\omega_p)}
\label{Eq:e-dispersion}
\end{multline}
where $n_o(\omega_p)$ and $n_e(\omega_p)$ are ordinary and extraordinary refractive indices of the crystal.
For each plane wave component of pump field with given $\kappa_p=(\omega_p,k_{px},k_{py})$ the above relation fixes
the $z$-component of the wavevector $k_{pz}$ at a value $k_{pz}(\kappa_p)$. During
propagation along $z$ each of the plane-wave components will acquire phase equal to $\exp[ik_{pz}(\kappa_p)L]$.
Thus, the equation of evolution of the pump pulse $\tilde A_p(\kappa_p,z)$ can be written as \cite{BandPRL96}:
\begin{align}\label{Eq:dAp/dz}
\frac{\partial}{\partial z}\tilde A_p(\kappa_p,z)=ik_{z,p}(\kappa_p)\tilde A_p(\kappa_p,z).
\end{align}

On the contrary, the parametric fluorescence has to be described by a quantum field.
We expand this field in a basis of plane waves parametrized by $\kappa=(\omega,k_{x},k_{y})$, which corresponds
to the following expression for the field operator $\hat E(x,y,z,t)$ in the position space:
\begin{multline}
\hat E(x,y,z,t)=\\
i\int \frac{d^3\kappa}{(2\pi)^3} \sqrt{\frac{\hbar\omega}{2\epsilon_0\sqrt{n(\omega)}}}
\hat a(\kappa,z) e^{ik_xx+ik_yy-i\omega t}  + \text{H.c.}
\label{Eq:hatE(x,y,z,t)}
\end{multline}
Here $\hat a(\kappa,z)$ form a complete set of bosonic annihilation operators at each $z$
while $n(\omega)$ is the index of refraction of the medium.
The evolution of the parametric fluorescence is also largely determined by the crystal dispersion,
which determines $k_z$ for each plane wave component with given $\kappa=(\omega,k_{x},k_{y})$
at a value $k_z=k_z(\kappa)$,
\begin{equation}\label{Eq:kz}
  k_z(\kappa)=\sqrt{n_o^2\frac{\omega^2}{c^2}-k_x^2-k_y^2}.
\end{equation}
Additionally, the signal field interacts with the pump pulse via the second order nonlinearity.
Classical equations, describing propagation of the fluorescence field through the crystal
are well known \cite{BandPRA94}. Quantization of the signal field corresponds to replacement
of c-number amplitudes and their complex conjugates of plane-wave modes by annihilation and creation
operators $\hat a(\kappa,z)$ and $\hat a^\dagger(\kappa,z)$ and
leads to the Heisenberg equations for their evolution:
\begin{multline}\label{Eq:da/dz}
\frac{\partial}{\partial z} \hat a(\kappa,z)=ik_z(\kappa) \hat a(\kappa,z) \\
+\frac{1}{\LNL} \int d\kappa'\frac{\tilde A_p(\kappa',z)}{A_0} \hat a^\dagger(\kappa'-\kappa,z).
\end{multline}
Here, $A_0$ is the maximal amplitude of the pump pulse in time-position space
$A_0=\int d^3\kappa_p \tilde A_p(\kappa_p)$,
while $\LNL$ is a nonlinear length, a constant which determines a characteristic length over which
nonlinear interaction produces a few photon pairs, defined as:
\begin{align}\label{Eq:LNL=}
  \frac{1}{\LNL}=\frac{\omega_p^2d_{\mathrm{eff}}A_0}{8c^2k(\omega_p/2)}.
\end{align}

For a crystal of a length $L$ and given pump pulses $\tilde A_p(\kappa,z=0)$ at the crystal entrance
we can find Green functions $C(\kappa,\kappa')$ and $S(\kappa,\kappa')$ of the
Eq.~\eqref{Eq:da/dz} and express the fluorescence field at the crystal exit
face $\hat a\out(\kappa)=\hat a(\kappa,L)$ in terms of that on the input face $\hat a\inp(\kappa)=\hat a(\kappa,0)$:
\begin{align}\label{Eq:aout=C+S}
  \hat a\out(\kappa)=\int d\kappa'
  \left[C(\kappa,\kappa') \hat a\inp(\kappa') + S(\kappa,\kappa') \hat a^\dagger\inp(\kappa')\right].
\end{align}

The above equation gives a complete quantum description of the parametric amplifier. In our case
it is  highly redundant, since we are interested in the average intensity of the parametric fluorescence
$\avg{\hat a^\dagger(\kappa)\hat a(\kappa)}=\avg{\hat n(\kappa)}$ as a function of the frequency and direction,
both contained in $\kappa$. In the following section, we use
 Eq.~\eqref{Eq:aout=C+S} only as a backbone for developing more specialized results.

\section{Perturbative approximation}
In this section we discuss the properties of the average intensity of the parametric fluorescence calculated in
the first order approximation with respect to the strength of the nonlinearity $1/\LNL$.
This approximation physically corresponds to a situation when the fluorescence field consists of a single
or a few pairs of photons emitted into distinct frequency or angular regions.
It is expected that as soon as the pump becomes so intense that the photons are emitted in bunches,
perturbative expansion fails. However, as we show, some results inferred from the perturbative approach, 
can be directly applied even in the high-gain regime. 

Basic expressions derived in this section are well known and have been introduced
in works on spontaneous parametric down-conversion in the context of photon sources \cite{GricePRA01}.
However, the intensity of such sources have, surprisingly,
received very little theoretical attention.

In order to calculate $\bar n(\kappa)=\avg{\hat a^\dagger(\kappa)\hat a(\kappa)}$ which describes the intensity of 
the parametric
fluorescence, we first obtain an approximate Green functions $C(\kappa,\kappa')$ and $S(\kappa,\kappa)$. This is done
by transforming into the interaction picture, which is accomplished by substituting:
\begin{align}\label{a=ikz aI}
\hat a(\kappa,z)=\exp[ik_z(\kappa)z]\hat a_I(\kappa,z),
\end{align}
into Eq.(\ref{Eq:da/dz}). The resulting equation of evolution of $\hat a_I(\kappa,z)$ contains only
nonlinear term and can be easily solved perturbatively with respect to $1/\LNL$ \cite{WasilewskiPRA06A}.
The first order solution is next compared with a general form given
by Eq.~\eqref{Eq:aout=C+S}. The $C(\kappa,\kappa')$ function in the lowest order describes simple dispersive evolution
of the fluorescence filed passing through the crystal:
\begin{equation}
  C(\kappa,\kappa')=\delta(\kappa-\kappa')e^{i k_z(\kappa) L} \label{Eq:C=1},
\end{equation}
while the $S(\kappa,\kappa')$ describes the process of pair generation:
\begin{multline}
  S(\kappa,\kappa')=\frac{L \tilde A_p(\kappa+\kappa')}{\LNL A_0}
  e^{i[k_z(\kappa)-k_z(\kappa')]\frac{L}{2}}\\
  \times\mathrm{sinc}\,
  \left(\frac{L\Delta k(\kappa,\kappa')}{2}\right),\label{Eq:S=sinc}
\end{multline}
where $\Delta k(\kappa,\kappa')$ is a wavevector mismatch between plane-wave components of the fluorescence characterized
by spatio-temporal frequencies $\kappa$ and $\kappa'$
and coupled plane wave component of the pump characterized by spatio-temporal frequency $\kappa+\kappa'$:
\begin{eqnarray}\label{Eq:Dk=kp-ks-ki}
  \Delta k(\kappa,\kappa')&=& k_{pz}(\kappa+\kappa')-k_{z}(\kappa)-k_{z}(\kappa')
\end{eqnarray}

The above result can be used to calculate the intensity of spontaneous parametric fluorescence
emitted at a certain direction in a certain frequency, corresponding to a vector $\kappa$:
\begin{multline}
  \bar n(\kappa)=\avg{\hat a\out^\dagger(\kappa)\hat a\out(\kappa)}=\int d^3\kappa'\ |S(\kappa,\kappa')|^2
  =\\
  \left(\frac{L}{\LNL}\right)^2\int d^3\kappa'\ \left|\frac{\tilde A_p(\kappa+\kappa')}{A_0}\right|^2\
  \mathrm{sinc^2}\left(\frac{L\Delta k(\kappa,\kappa')}{2}\right).\label{Eq:n=int}
\end{multline}
\begin{figure}
 	\centering
	\includegraphics[scale=1.2]{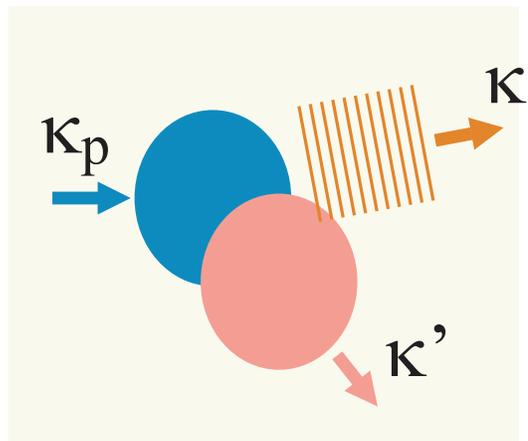}
 	\caption{(color online). Observation of a plane-wave monochromatic component of the fluorescence field,
 	  directed to the upper right in the picture,
 	  leads to projection of the conjugate idler field into a pulsed form, directed to the lower right.
 	The idler wave-packet takes up all the frequency and wavevector
 	span of the pump pulse, incoming from the left in the picture.}
 	\label{fig:3WM}
\end{figure}
The above integral can be given a meaningful physical interpretation. 
Observation of a signal photon testifies generation of the idler photon. In our case
the first is detected using narrow-band filters and is 
nearly plane-wave and monochromatic, characterized by vector $\kappa$. Thus,
all the frequency and direction distribution present in the pump is transferred 
to the idler photon, as depicted in Fig.~\ref{fig:3WM}. Each plane-wave component of the idler characterized by $\kappa'$
can support the production of signal characterized by $\kappa$ if the pair in question
is phase-matched with a pump component characterized by $\kappa+\kappa'$, i.e. the 
phase mismatch $L\Delta k(\kappa,\kappa')$ is small.

The  integral given in Eq.~\eqref{Eq:n=int} cannot be evaluated analytically in case of pulsed pumping,
when $\tilde A_p(\kappa_p,z)$ is nonzero over a finite range of $\kappa$.
However, the integrand can be approximated with a Gaussian function.
First, let us assume a Gaussian form of the pump amplitude at the crystal entrance face $\tilde A_p(\kappa_p,z=0)$, 
corresponding to a pulse of duration $\tau_p$
propagating in a beam of width $w_p$, centered around frequency $2\omega_0$:
\begin{multline}\label{A_p=gauss}
  \tilde A_p(\kappa_p,z=0)=
  A_0 \frac{w_p^2 \tau_p}{(2\pi)^{3/2}}\\
  \times\exp\left(-\frac{\tau_p^2(\omega-2\omega_0)^2}{2}-\frac{w_p^2(k_x^2+k_y^2)}{2}\right).
\end{multline}

\begin{figure}
  \center\includegraphics[scale=0.8]{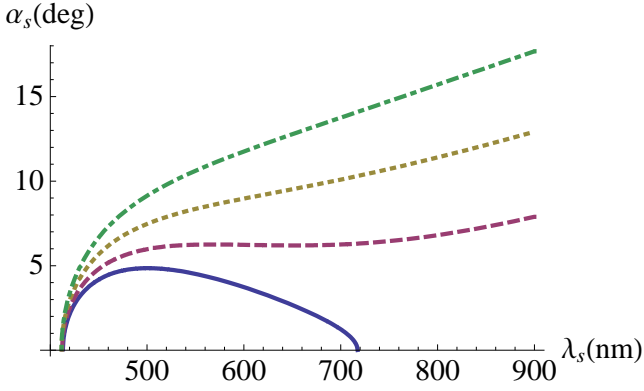}
  \caption{(color online). Angle $\alpha_s=\arcsin[ck^0(\omega)/\omega]$ at which perfectly phase-matched fluorescence at
    the wavelength $\lambda=2\pi c/\omega$ emerges outside the crystal.
    Plot corresponds to type-I phase-matching in BBO pumped
    at $2\pi c/(2\omega_0)=400\,$nm for crystals cut at angles $\theta=29^o$ (solid line),
    $31.3^o$ (dashed line), $35^o$ (dotted line) and $40^o$ (dash-dot line). }\label{fig:tana_vs_ls}
\end{figure}

Concentration of the pump energy around given $\kappa_p$
together with requirement of phase matching $\Delta k(\kappa,\kappa')<1/L$ reduces the range of
spatio-temporal frequencies $\kappa$ and $\kappa'$ over which
a significant amount of parametric fluorescence is generated.
To analyze this effect, let us consider perfectly phase-matched parametric down-conversion of the central
plane-wave component of the pump with $\kappa_p=(2\omega_0,0,0)$. This part of the fluorescence 
consists of a plane wave component-pairs characterized by $\kappa$ and $\kappa'$ which
satisfy the following set of equations:
\begin{align}\label{Eq:Ap=max,Dk=0}
\left\{\begin{matrix}
\kappa+\kappa'=(2\omega_0,0,0)\\
\Delta k(\kappa,\kappa')=0.
\end{matrix}\right.
\end{align}
The first equation corresponds to a requirement that both components of the down-conversion couple to the
central component of the pump, while the second equation is the perfect phase matching condition.
This is in fact a set of four equations for six components of $\kappa$ and $\kappa'$ which defines a two-dimensional surface.
Assuming $\kappa=(\omega,k_x,k_y)$ as before, from the first equation we get $\kappa'=(2\omega_0-\omega,-k_x,-k_y)$. 
Then the solution of the second equation can be cast into the following form:
\begin{equation}
\sqrt{k_x^2+k_y^2}=k^0(\omega).
\end{equation}
This is an axially-symmetric surface in three-dimensional space of $\kappa=(\omega,k_x,k_y)$,
describing plane-wave components of the parametric fluorescence which can be produced from the central component
of the pump in a perfectly phase-matched process.  In Fig.~\ref{fig:tana_vs_ls} we plot a section through 
this surface in the
$k_x$--$\omega$ plane, for parametric down conversion in a BBO crystal pumped at $2\pi c/(2\omega_0)=400\,$nm and
four different cut angles $\theta$.

As the spatio-temporal frequency of the fluorescence
$\kappa$ deviates from the perfect phase matching surface $k^0(\omega)$ its intensity diminishes.
In the region of $\kappa$ in which the fluorescence intensity is large, the sinc function
can be approximated by a Gaussian \cite{Dragan2004}. Such an approximation fails only for
very thick crystals, in which intricacies of the phase matching may play important role, but
the fluorescence emitted in such cases contains very complicated correlations, which render it useless
in most of the known applications.
\newcommand{\obs}{\mathrm{obs}}

The next step towards Gaussian approximation of the integrand in Eq.~\eqref{Eq:n=int} is a linearization of the phase mismatch $\Delta k(\kappa,\kappa')$
given by Eq.~\eqref{Eq:Dk=kp-ks-ki} around points on perfect
phase-matching surface. Since it is axially symmetric, we choose points in $k_x>0$ and $\omega$ half-plane.
For each fixed observation frequency $\omega_\obs$ we expand phase mismatch $\Delta k(\kappa,\kappa')$ around
$\kappa_0=(\omega_\obs,k^0(\omega),0)$ and $\kappa'_0=(2\omega_0-\omega_\obs,-k^0(\omega),0)$ up to linear terms. 
This gives
\begin{multline}
  \Delta k_z^{\mathrm{lin}}(\kappa,\kappa')=
  (2\omega_0-\omega'-\omega_\obs)\Delta\beta'_1
  +(k_x-k^0(\omega_\obs))\Delta\rho_{x}\\
  +k_y\Delta\rho_{y}+
  (k'_x+k^0(\omega_\obs))\Delta\rho'_{x}+k'_y\Delta\rho'_{y},
  \label{Eq:Dklin}
\end{multline}
where the expansion coefficients equal:
\begin{eqnarray}
\Delta\beta'_1&=&\frac{\partial k_{pz}(\kappa_p)}{\partial\omega}\Bigr|_{\kappa_p=\kappa_0+\kappa'_0}-
  \frac{\partial k_z(\kappa')}{\partial\omega'}\Bigr|_{\kappa'=\kappa'_0},\nonumber\\
\Delta\rho_{i}&=&\frac{\partial k_{pz}(\kappa_p)}{\partial k_{pi}}\Bigr|_{\kappa_p=\kappa_0+\kappa'_0}-
  \left.\frac{\partial k_z(\kappa)}{\partial k_{i}}\right|_{\kappa=\kappa_0}\nonumber\\
\Delta\rho'_{i}&=&\frac{\partial k_{pz}(\kappa_p)}{\partial k_{pi}}\Bigr|_{\kappa_p=\kappa_0+\kappa'_0}-
  \left.\frac{\partial k_z(\kappa')}{\partial k'_{i}}\right|_{\kappa'=\kappa'_0},\nonumber \\
 &&\qquad i=x,y. \label{Eqs:coeffs}
\end{eqnarray}
The above coefficients have simple physical meaning. $\Delta\beta'_1$ is a difference of inverse group velocities 
between the pump pulse
and the fluorescence component characterized by $\kappa'$, while $\Delta\rho$ give the amount of spatial displacement 
per unit length
between pump pulses and the fluorescence component characterized by $\kappa$ or $\kappa'$ in $x$ or $y$ direction.
Note, that all the above coefficients depend on the choice of the observation frequency $\omega_\obs$.

Finally, in Eq.~\eqref{Eq:n=int} we approximate the sinc function by a Gaussian
\begin{equation}
\sinc^2\left(\frac{L\Delta k_z}{2}\right)
\simeq\exp\left(-\frac{L^2}{12}(\Delta k_z^{\mathrm{lin}})^2\right)
\label{Eq:sinc=gauss}
\end{equation}
and obtain the following expression for the fluorescence intensity
\begin{multline}
\bar n(\kappa)=\left(\frac{L}{\LNL}\right)^2 \\
  \times\int d^3\kappa'
  e^{-\frac{L^2}{12}[\Delta k_z^{\mathrm{lin}}(\kappa,\kappa')]^2} \left|\frac{\tilde A_p(\kappa+\kappa')}{A_0}\right|^2.
  \label{Eq:n=intlin}
\end{multline}
This integral can be evaluated analytically. Let us substitute $\kappa=\kappa_0=(\omega,k^0(\omega),0)$
which describes the perfect phase-matching line. Then we obtain an estimate of the intensity of the 
fluorescence emitted at a frequency $\omega$ at an angle
$\alpha=\arctan(ck^0(\omega)/\omega)$ to the $z$ axis in $z-x$ plane:
\begin{multline}
  n(\omega_\obs,k^0(\omega_\obs),0)=\frac{w_p^2\tau}{4\pi^{3/2}}
  \left(\frac{L}{\LNL}\right)^2 \\
  \times\left[4 +\frac{L^2(\Delta\rho'^2_{x}+\Delta\rho'^2_{y})}{w^2_p}
    +\left(\frac{L\Delta\beta'_1}{\tau_p}\right)^2
    \right]^{-1/2}.\label{Eq:n=1/D}
\end{multline}
Let us consider the dependence of the intensity of the fluorescence along the perfect phase-matching line
$n(\omega_\obs,k^0(\omega_\obs),0)$ on the frequency $\omega_\obs$ at which we observe it. Note again, that observation of
the fluorescence signal at the frequency $\omega_\obs$ testifies generation of the conjugate (idler) photons at the
frequency $2\omega_0-\omega_\obs$. It turns out, that the propagation properties of the idler photons influence the
intensity of the signal. In particular three parameters, which vary with $\omega_\obs$ come into play. The first is the
ratio of the temporal walk-off between pump pulse and conjugate photons over the crystal length $L\Delta\beta'_1$ to the
pump pulse duration $\tau_p$. The other two measure the ratio of the transverse displacement between the  pump pulse
and conjugate photons $L\Delta\rho'_x$ and $L\Delta\rho'_y$ to the pump pulse width $w_p$.

This can be understood by reconsidering picture drawn in Fig.~\ref{fig:3WM}. As we argued before, observation of a nearly
plane-wave signal transfers the frequency and wavevector distribution of the pump to the idler photon,
which takes a form of a short pulse and propagates according to the crystal dispersion. Usually, it quickly diverges 
from the pump pulse --- the expression in the square brackets in Eq.~\eqref{Eq:n=1/D} is large. 
However, for some frequencies of the signal $\omega_\obs$ the idler propagates with the pump for 
a longer distance which increases the nonlinear interaction and the photon flux $n(\omega_\obs,k^0(\omega_\obs),0)$.

\begin{figure}\center
  \includegraphics[scale=0.82]{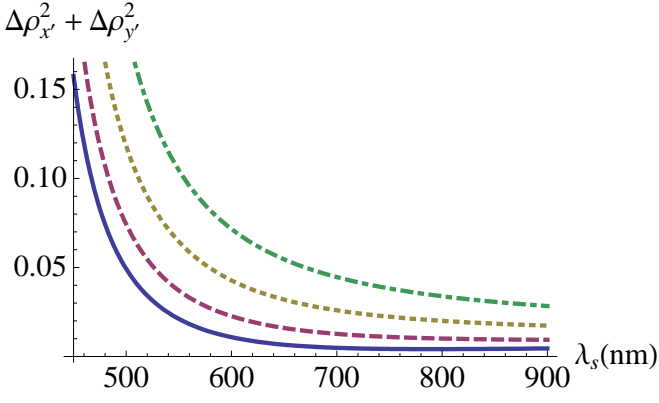}
  \caption{(color online). Square idler spatial displacement coefficient $\Delta\rho_{x'}^2+\Delta\rho_{y'}^2$ in BBO for 
    four different crystal cut angles $\theta=29^\circ$ (solid line), $\theta=31.3^\circ$ (dashed line), 
    $\theta=35^\circ$ (dotted line) and  $\theta=40^\circ$ (dot-dashed line).}\label{fig:rho}
\end{figure}

\begin{figure}\center
  \includegraphics[scale=0.7]{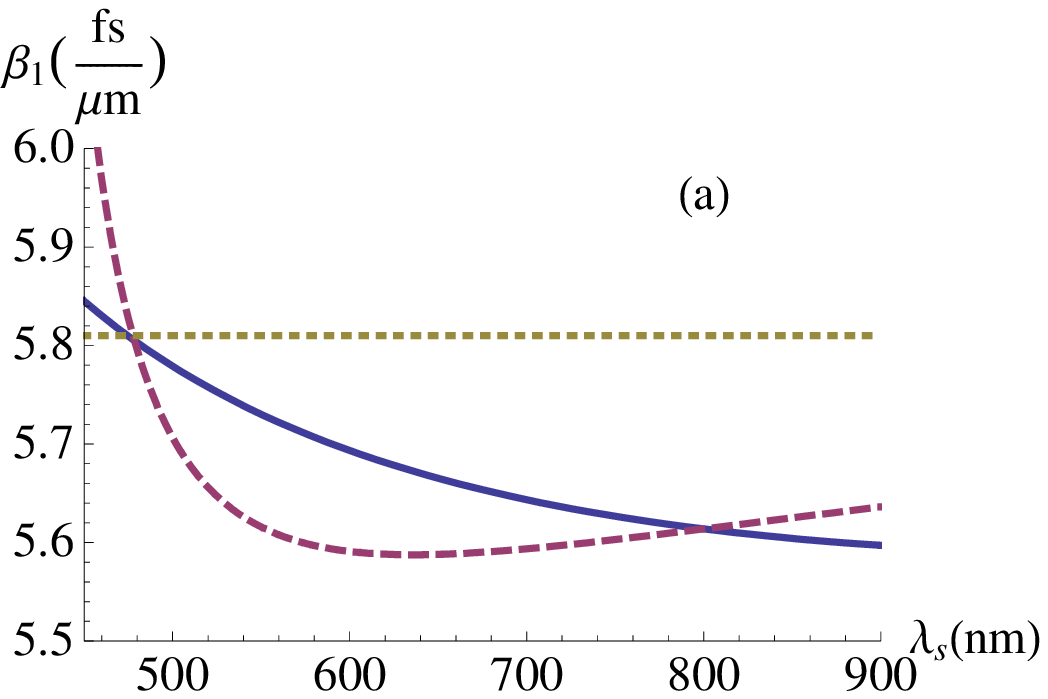}
  \includegraphics[scale=0.7]{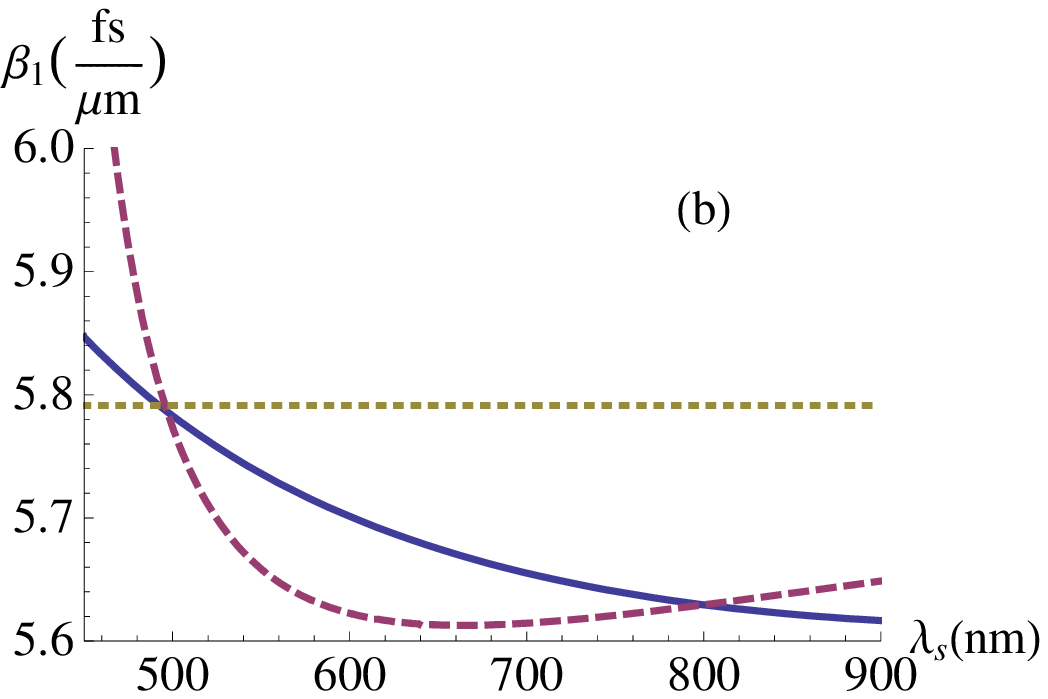}
  \includegraphics[scale=0.7]{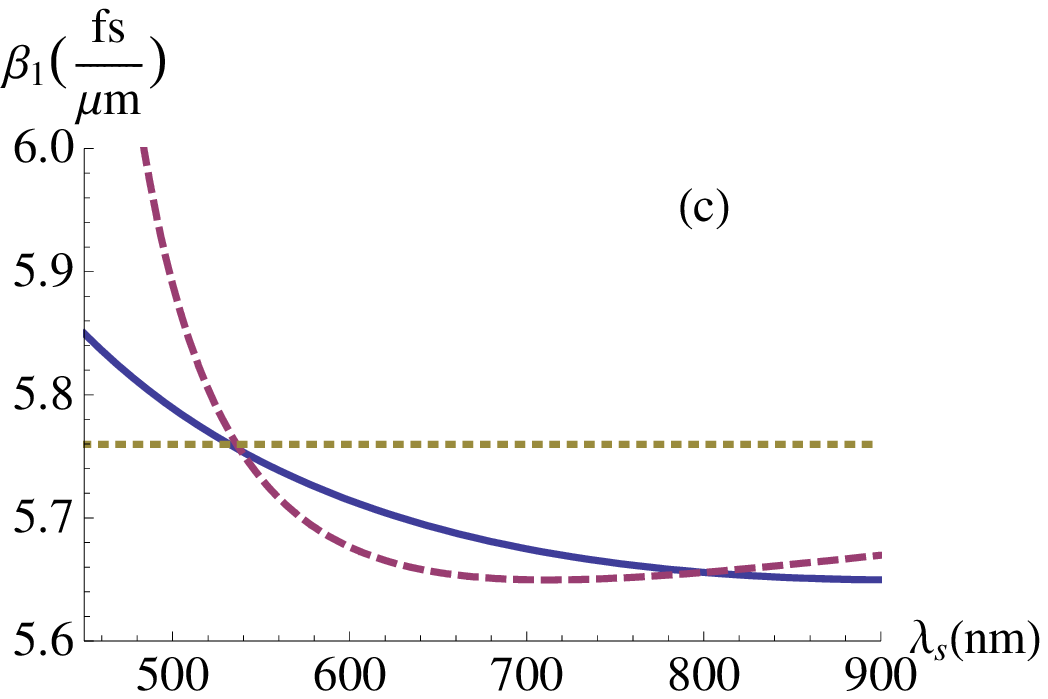}
  \includegraphics[scale=0.7]{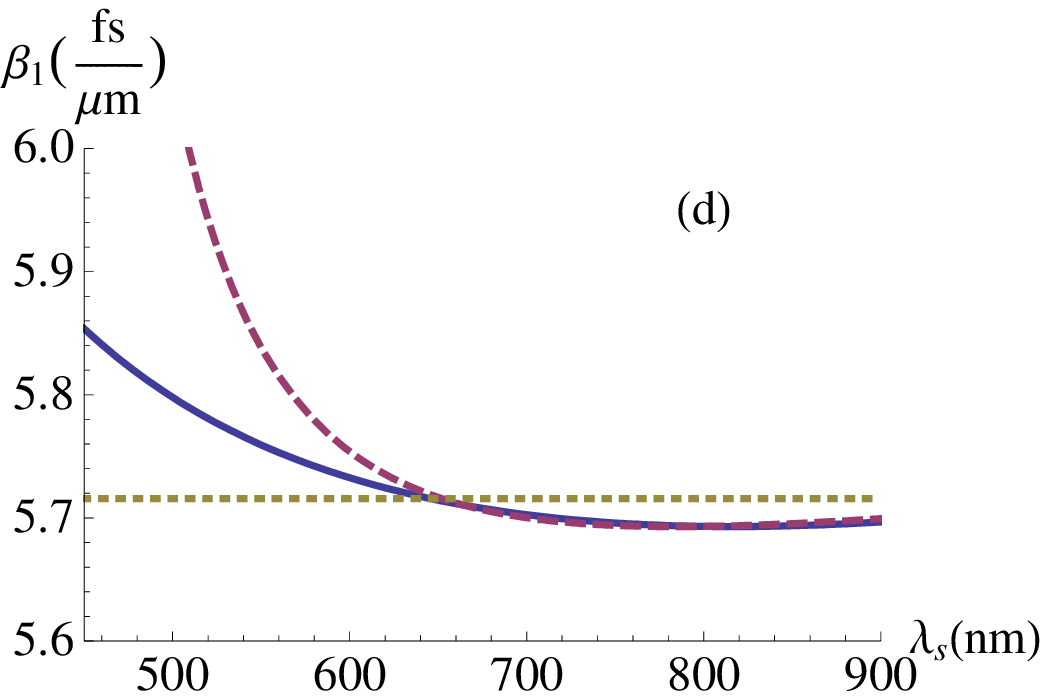}
  \caption{(color online).
    Inverse of the group velocities of pump (dotted line), signal (solid line) and idler (dashed line)
  for following angles of the crystal setting:
  (a) $\theta=29^o$, (b) $\theta=31.3^o$, (c) $35^o$ and (d) $40^o$. }\label{fig:beta1}
\end{figure}

We plot relevant material coefficients in BBO in case
when it is pumped at $400$nm and cut at four characteristic crystal cut angles $\theta$ 
in Fig.~\ref{fig:rho} and \ref{fig:beta1}.

\begin{figure}\center
  \includegraphics[scale=0.7]{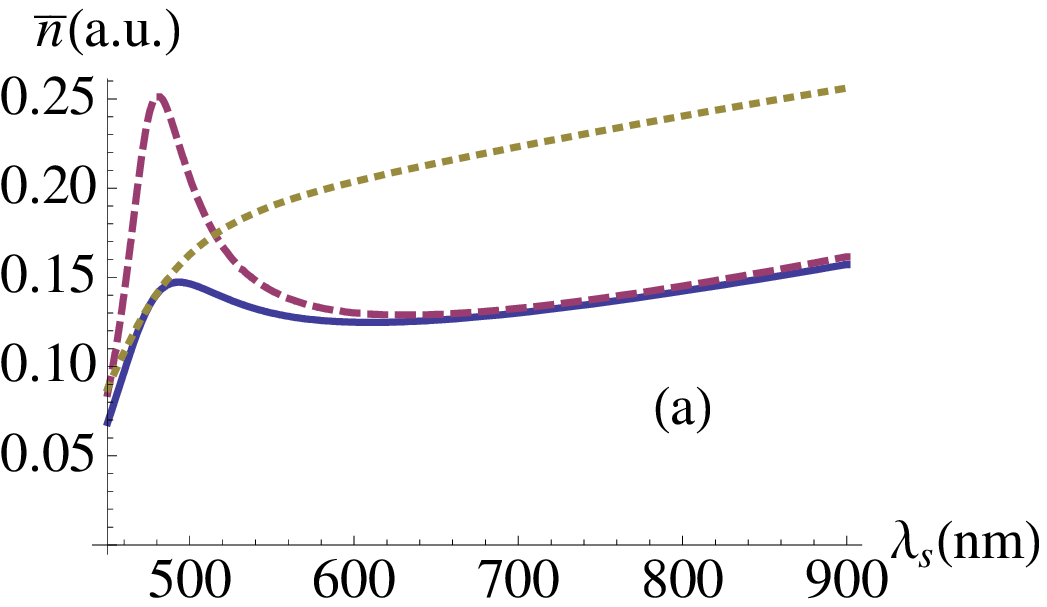}
  \includegraphics[scale=0.7]{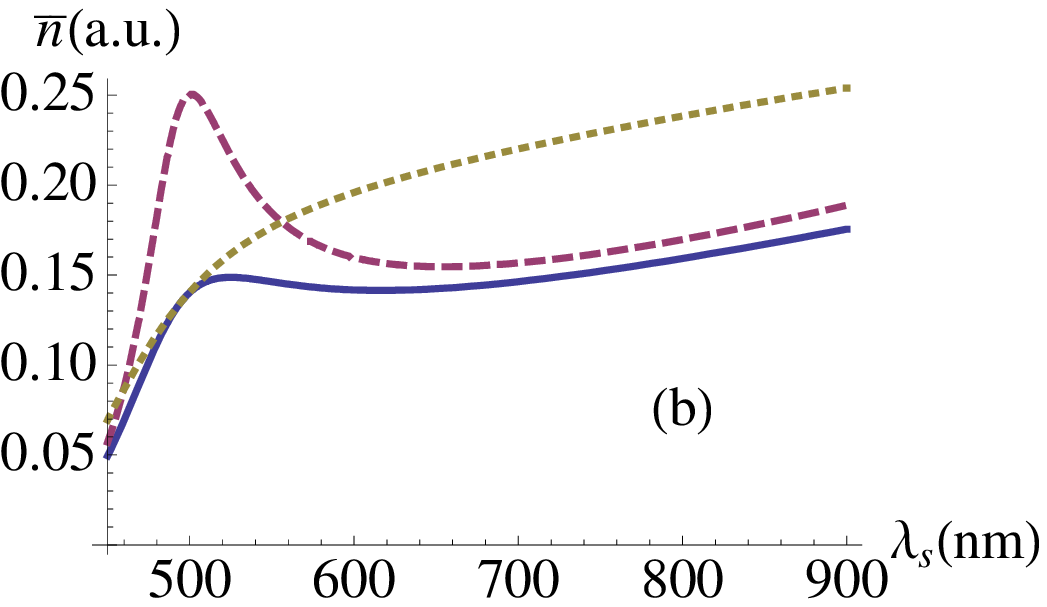}
  \includegraphics[scale=0.7]{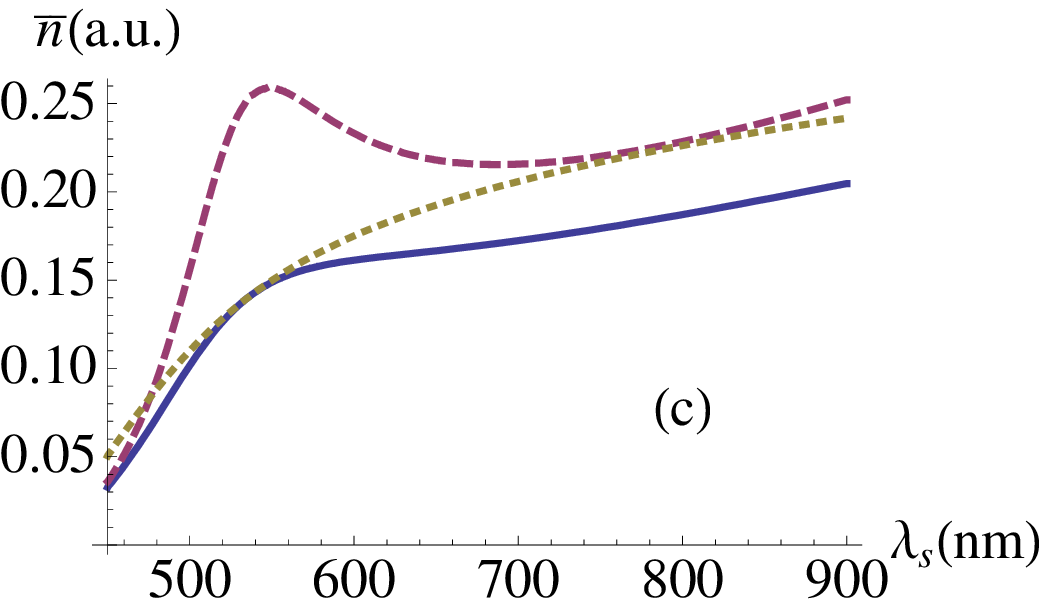}
  \includegraphics[scale=0.7]{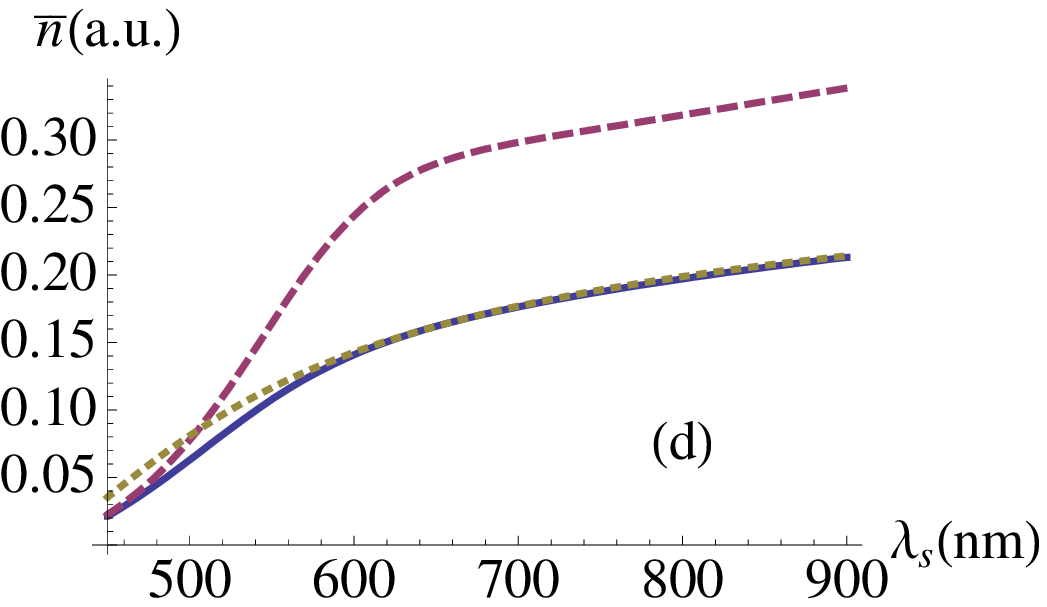}
  \caption{(color online). Photon flux of the fluorescence $\bar n(\omega_\obs,k^0(\omega_\obs),0)$ 
  along the perfect phase matching line plotted in Fig.~\ref{fig:tana_vs_ls}
   calculated using Gaussian approximation for 2mm long BBO crystal.
  Plots are made for four different crystal cut angles: (a)
  $\theta=29^o$, (b) $31.3^o$, (c) $35^o$ and (d) $40^o$. Various lines indicate different pump pulse parameters:
 	solid lines correspond to $\tau_p=$60 fs, $w_p=$80 $\mu$m, 
  dashed lines correspond to larger pump beam, $\tau_p=$60 fs, $w_p=$160 $\mu$m 
  and dotted lines correspond to longer pump pulses, $\tau_p=$120 fs, $w_p=$80 $\mu$m.}\label{fig:intensity}
\end{figure}

We also plot the intensity of the fluorescence along the perfect phase-matching line
$n(\omega_\obs,k^0(\omega_\obs),0)$ given in Eq.~\eqref{Eq:n=1/D}
as a function of the wavelength of observation $2\pi c/\omega_\obs$
in Fig.~\ref{fig:intensity} for a few typical pump pulse durations and widths.

It is worthwhile to consider the features displayed by this plots.
For short pump pulses, when temporal walk-off between pump and conjugate photons is more pronounced,
the fluorescence intensity reaches local maximum values at a wavelength where this walk-off vanishes $\Delta\beta_1=0$.
For longer pulses the temporal walk-off loses its importance, and the intensity of the fluorescence increases
with wavelength due to reduced spatial displacement between the pump and conjugate photons, i.e.
$\Delta\rho'^2_{x}+\Delta\rho'^2_{y}$.
\section{Three dimensional stochastic model}
In the previous section we calculated the intensity of the parametric fluorescence 
in the single pair generation limit. Now we switch to intense-pumping regime, when photons are generated in bunches 
and perturbative approximation breaks down. 
Estimation of the fluorescence intensity is 
of a great importance in novel practical applications of parametric amplifiers. For example, when using 
this kind of nonlinear interaction to amplify broadband chirped pulses, the parametric fluorescence is a 
source of large and difficult to avoid noise \cite{TavellaNJP06}. 

In this section we describe a numerical method of solving the dynamics of a coupled pump-signal system
given by Eqs.(\ref{Eq:dAp/dz}) and (\ref{Eq:da/dz}) and estimating the intensity of the parametric fluorescence. 
We use so-called stochastic Wigner method, which can in principle provide exact solutions for linear quantum systems,
such as described by Eq.~\eqref{Eq:da/dz}.
If the Wigner function of the field at the entrance is Gaussian, in particular representing vacuum state, the
Wigner function of the final state of the parametric fluorescence will be Gaussian too. It is easily deduced
from a general form of the input-output relations given by Eqs.(\ref{Eq:aout=C+S}).

A crucial observation is that if the Heisenberg equation is linear, the evolution
of the Wigner function is the same as a classical Liouville equation for the density
probability of finding the system in a state with particular electric field. Thus the Wigner function
$W[\alpha(\kappa,z)]$ can be interpreted as a probability density of finding the system with a fluorescence
field
described by a spectral amplitude $\alpha(\kappa,z)$.

\begin{figure}
  \centering
  \includegraphics[scale=0.6]{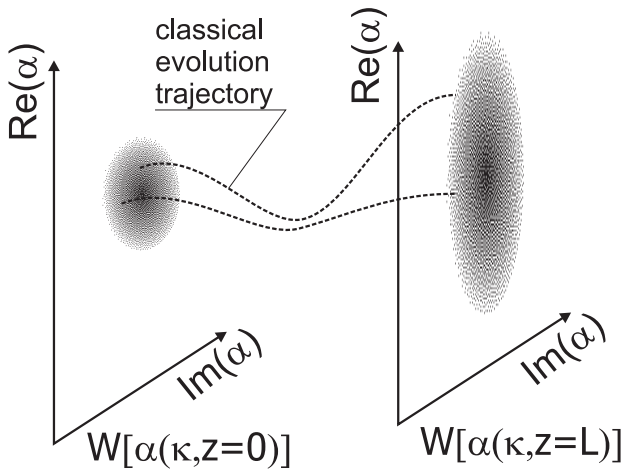}
  \caption{Graphical illustration of the Wigner method.}
  \label{fig:wigner}
\end{figure}

Therefore, to find the Wigner function of the fluorescence at the crystal exit face we 
repeat the following procedure.
First we draw an input spectral amplitude $\alpha(\kappa,z=0)$ with probability given
be the vacuum Wigner function
\begin{equation}
W[\alpha(\kappa,z=0)]=\left(\frac{2}{\pi}\right)^N\exp\left(-2\sum_\kappa |\alpha(\kappa)|^2\right),
\label{Eq:Wvacuum}
\end{equation}
where we assumed a finite number $N$ of discrete modes indexed by $\kappa$.
Next we solve classical equations of propagation of the signal field:
\begin{equation}\label{Eq:classical}
  \frac{\partial}{\partial z}\alpha(\kappa,z)=ik_z(\kappa) \alpha(\kappa,z)
  +\int d\kappa' \frac{\tilde A_p(\kappa,z) \alpha^*(\kappa-\kappa',z)}{\LNL A_0}.
\end{equation}
This way we obtain a possible output spectral amplitude $\alpha(\kappa,z=L)$. By repeating this procedure
many times, we obtain a set of probable output field amplitudes $\left\{\alpha(\kappa,z=L)\right\}$,
which approximates Wigner function $W[\alpha(\kappa,z=L)]$. This is schematically illustrated in Fig.~\ref{fig:wigner}.

This can be used to calculate the mean photon flux
at a given direction and frequency. It equals a mean $|\alpha(\kappa)|^2$
averaged with the Wigner function $W[\alpha(\kappa,z=L)]$ minus vacuum average.
Using the set of output amplitudes $\left\{\alpha(\kappa,z=L)\right\}$ this is calculated as: 
\begin{equation}
\bar n(\kappa)=\left(\sum_{ \alpha(\kappa) \in \left\{\alpha(\kappa,z=L)\right\} } 
|\alpha(\kappa)|^2 \right)
  -\frac{1}{2}. \label{Eq:n=W.|a|^2}
\end{equation}

\begin{figure*}
  \centering
  \begin{tabular}{ccc}
  \includegraphics[scale=0.8]{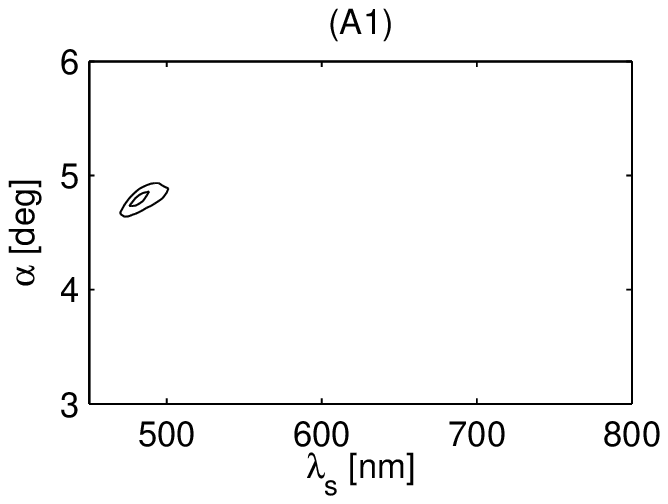}&
  \includegraphics[scale=0.8]{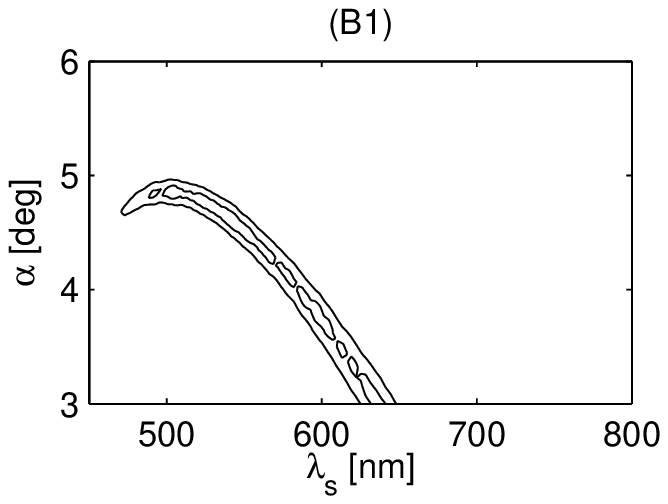}&
  \includegraphics[scale=0.8]{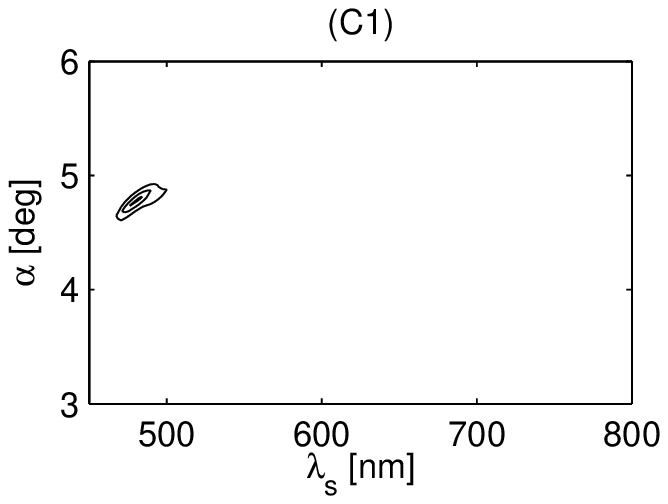}\\
  \includegraphics[scale=0.8]{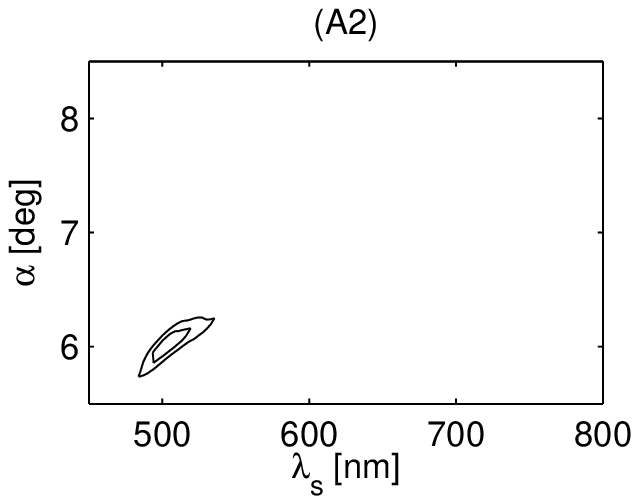}&
  \includegraphics[scale=0.8]{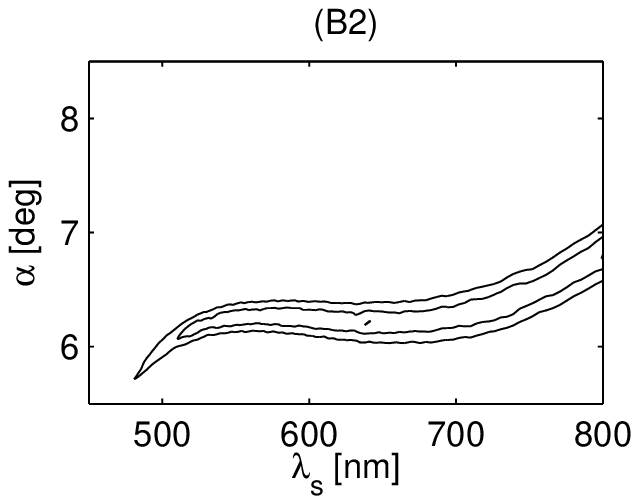}&
  \includegraphics[scale=0.8]{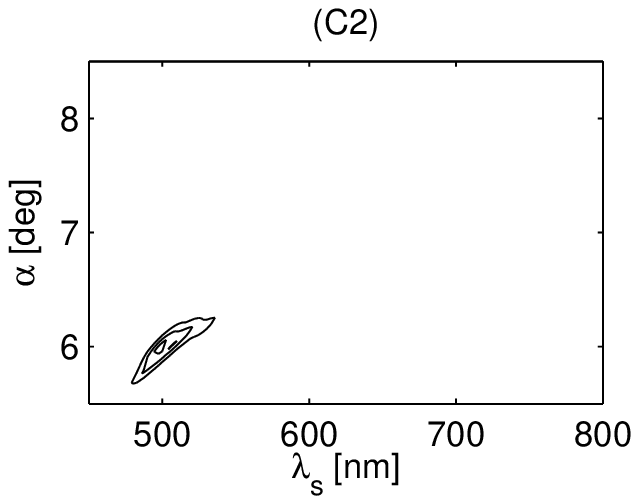}\\
  \includegraphics[scale=0.8]{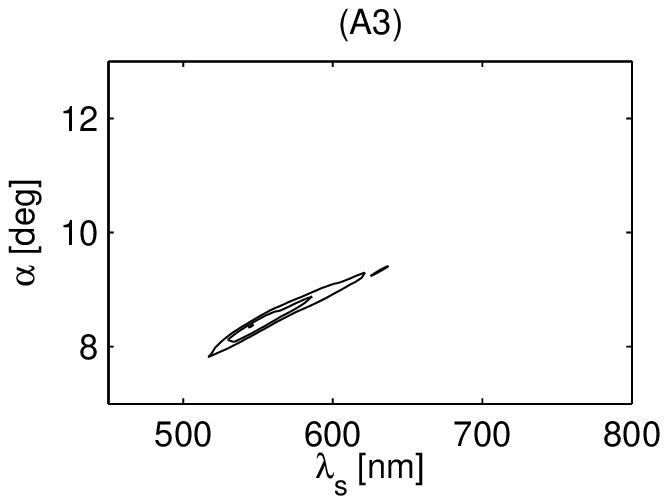}&
  \includegraphics[scale=0.8]{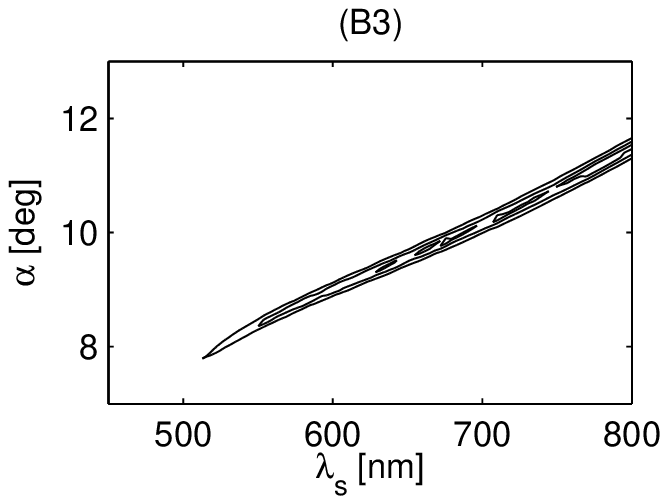}&
  \includegraphics[scale=0.8]{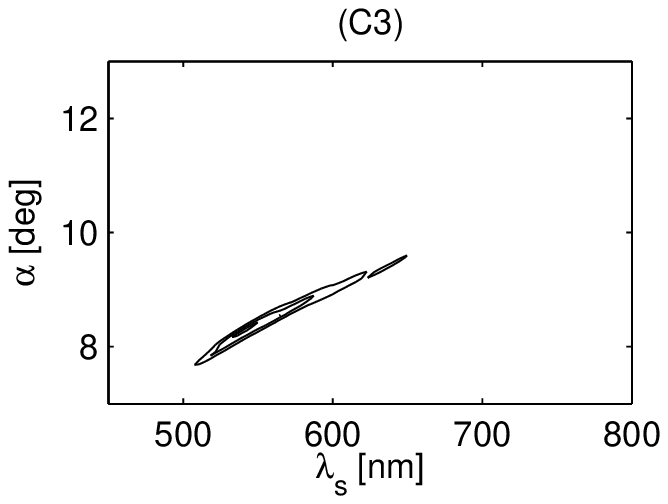}\\
  \includegraphics[scale=0.8]{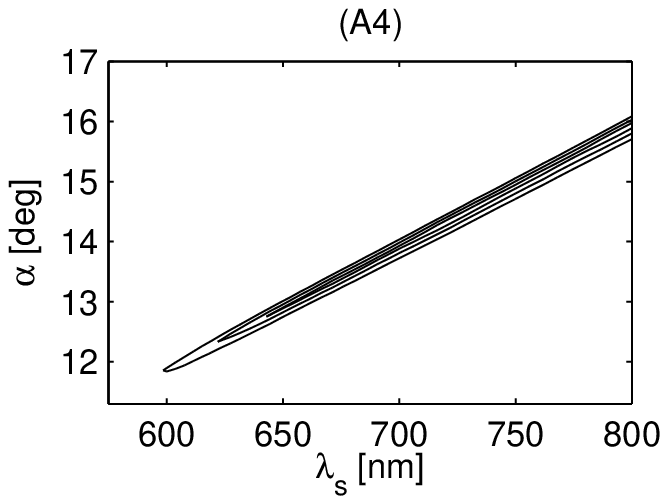}&
  \includegraphics[scale=0.8]{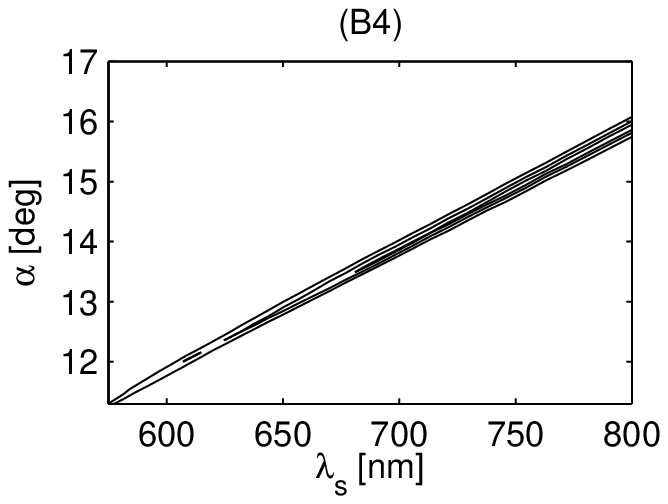}&
  \includegraphics[scale=0.8]{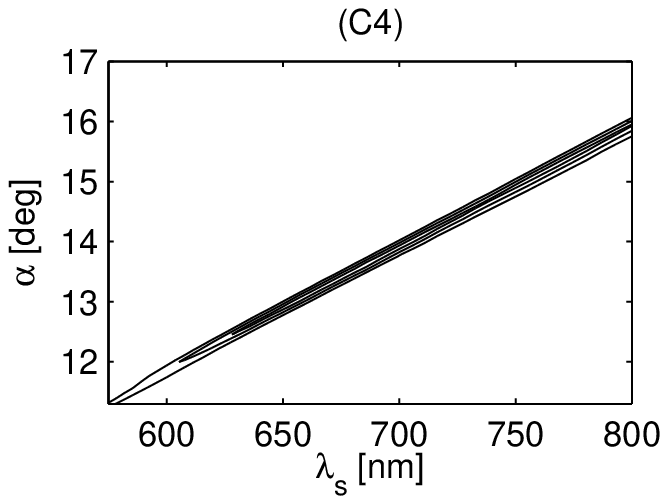}
  \end{tabular}
  \caption{
    The contour plot of the averaged photon flux of the fluorescence as a function of
    the angle $\alpha=\arcsin(c|k_x|/\omega)$ and
    the wavelength $\lambda=2\pi c/\omega$ for a 2mm long BBO. Each row corresponds to
    different crystal cuts:  $\theta=29^\circ$ (1st row), $\theta=31.3^\circ$ (2nd row), $\theta=35^\circ$ 
    (3rd row) and $\theta=40^\circ$ (4th row).
    Each column corresponds to different pump parameters: 
    column A is for $\tau_p=$60 fs, $w_p=$80 $\mu$m, 
    column B is for $\tau_p=$60 fs, $w_p=$160 $\mu$m 
    and column C is for  $\tau_p=$120 fs, $w_p=$80 $\mu$m. The contour lines were drawn at 0.25, 0.5
    and 0.75 of the maximum of each plot. 
  }

  \label{29}
\end{figure*}
We performed the procedure outlined above for interaction in 2mm long BBO crystal. 
The pump intensity was adjusted so that the total number of photons in the fluorescence 
field is near $10^8$. This corresponds to the typical conditions found in noncollinear-OPA (NOPA)
used for amplification of ultrashort pulses \cite{RiedleOL1997}.

For each of the various pump pulse durations, widths and crystal cut angles we performed ten independent
simulations of classical fluorescence evolution. The average photon flux was computed according 
to the formula \eqref{Eq:n=W.|a|^2}. Additionally, we assumed that $\bar n(\kappa)$
is almost independent of the angle around $z$ axis and we averaged the results over this angle. This
last step made ten independent simulations sufficient to calculate smooth intensities.

Figure \ref{29} shows the results of the simulations. 
The area of high intensity winds around the perfect phase-matching curve pictured in Fig.\ref{fig:tana_vs_ls},
as we argued before. 
Moreover, qualitatively, the intensity along this curve changes as predicted by the perturbation theory and
plotted in Fig.~\ref{fig:intensity}.
This can be considered an unexpected success of the perturbation theory and qualitatively explained as follows.
The process of parametric amplification with intense pumping leads to intense fluorescence fields, which exhibit mainly
classical properties. In particular 1/2 in Eq.~\eqref{Eq:n=W.|a|^2} can be neglected, which makes the
process virtually identical to a classical amplification of a weak, white noise. In the course of interaction
the components of the fluorescence become pairwise correlated. To some extend, the noise-like fluorescence
can be separated into pairs of peaks which have approximately opposite $k_x$ and $k_y$ and frequencies summing
up to the frequency of the pump, $2\omega_0$. Most of such pairs quickly diverge away from the pump pulse,
but some propagate with it enjoying exponential growth. This is very similar to the mechanism behind the
intensity of the fluorescence in a single pair generation regime, hence the intensity profiles are very similar.

It is worthwhile to compare the numerical results from Fig.~\ref{29} with  Figures \ref{fig:rho} and \ref{fig:beta1}. 
We find that for a short pump pulses
the temporal walk-off between the pump and the fluorescence must be very small to achieve intense generation. 
For longer pulses, this requirement is relaxed and our calculations reveal broadband parametric generation.
\section{Conclusions}
In this article we adopted perturbative and Wigner stochastic approaches to the task of calculating
an intensity of the parametric fluorescence. In a single pair generation limit, some analytical expressions
for the down-converted field
can be derived. By linearizing the phase-matching function we 
found simple formulas for the intensity of the fluorescence in this regime. Photon flux of the signal 
depends on the temporal and spatial walk-off between the idler and the pump. In case of type I interaction 
in BBO this leads to peaks in the spectrum of the downconversion signal at the frequencies where group velocity
of the idler matches up with the pump.

In an intense pumping case we performed a number of numerical simulations and found that the
spectrum of the fluorescence exhibits qualitatively the same features as predicted by the perturbative 
approximation. The contrast between the wavelengths at which the fluorescence is intense and those 
at which is weak is more apparent due to exponential character of amplification in non-perturbative regime.

The above results could be used as guidelines for constructing more bright photon-pair sources. They
could be also used for optimization of parametric amplifiers, in which one seeks minimal contribution from
the parametric fluorescence.

\section*{Acknowledgments}
We acknowledge insightful discussions with Konrad Banaszek, Marek Trippenbach and Czes{\l}aw Radzewicz,
as well as financial support from Polish Government scientific grant (2007-2009). 

\end{document}